\documentclass{article}

\usepackage[english]{babel}
\usepackage[letterpaper,top=2cm,bottom=2cm,left=3cm,right=3cm,marginparwidth=1.75cm]{geometry}

\usepackage{amsmath}
\usepackage{graphicx}
\usepackage[colorlinks=true, allcolors=blue]{hyperref}
\usepackage{subfigure}
\usepackage{float}

\title{Uplink of Visible Light Communication Enabled by\\ Receiving Ultrasonic Beamforming}

\author{Ming Che}
\date{}

\begin{document}
\maketitle

\begin{abstract}
We propose a method that uses inaudible ultrasonic waves to address the issue of uplink transmission in visible light communication. Our scheme, which applies frequency-shift keying modulation and receive beamforming by a microphone array, has been experimentally confirmed. This system has an adjustable receiving direction to enhance the anti-interference ability of uplink transmission, and it meets the asymmetric requirements of uplink and downlink service bandwidth for Internet services.
\end{abstract}

\section{Introduction}
The bi-directional transmission of visible light communication (VLC) has been a challenge for engineers for a long time. Without an effective uplink solution, the application scope of visible light communication will be limited, only allowing for one-way transmission of audio and video data or broadcast communication. Consequently, the successful implementation of a two-way visible light communication is necessary for a wireless internet experience, with the uplink being an essential aspect.

Komine et al. proposed a bi-directional transmission solution using a corner cube retro-reflector in 2003, however, this method was limited in terms of transmission rate and modulation bandwidth \cite{chap3_Komine_bidir}. Currently, there are many radio frequency based solutions such as WiFi, Bluetooth and ZigBee, however, these are not suitable to be used in places where radio frequency is prohibited, like hospitals and airplane cabins \cite{chap3_Shao_7035746,chap3_Shao_7293077,chap3_Yin_53643534,chap3_Kim_7423307}. Jaafar et al. proposed an infrared technology based solution for uplink transmission, however, this too has its own limitations such as low transmission rate and high directivity requirement \cite{chap3_Alresheedi_2017uplink,chap3_Alsulami_2019infrared}.

The use of all-optical bi-directional transmission technology utilizes LED as a signal source to enable full duplex communication. Two multiplexing technologies, time division duplexing (TDD) and frequency division duplexing (FDD), are commonly used to achieve this. TDD technology receives and transmits in different time slots and same frequency band, and has been used by Liu et al ~\cite{chap3_liu_2012demonstration}. to realize bidirectional transmission with OOK modulation, achieving a data transmission rate of 2.5 Mbit/s. Meanwhile, Wang et al. used three different wavelengths of RGB-LED for uplink and downlink data, with blue LED as the light source for the downlink and red and green LED as the light source for the uplink. This FDD-based system achieved an offline test throughput of 800 Mbit/s within 66 cm of free-space transmission \cite{chap3_wang_2013875,chap3_wang_2013demonstration}. While the uplink transmission scheme based on visible light is capable of higher transmission rates, its visible light characteristics limit its application scenarios since uplink does not require illumination.

This work focuses on overcoming a persistent challenge in VLC - the difficulty of effective uplink transmission. 
Traditional approaches have their limitations, so we have suggested non-audible audio signals and microphone arrays for data reception as alternative solutions.

\newpage

\section{Method and Demonstration}

\begin{figure}[H]
\centering
\includegraphics[width=0.49\columnwidth]{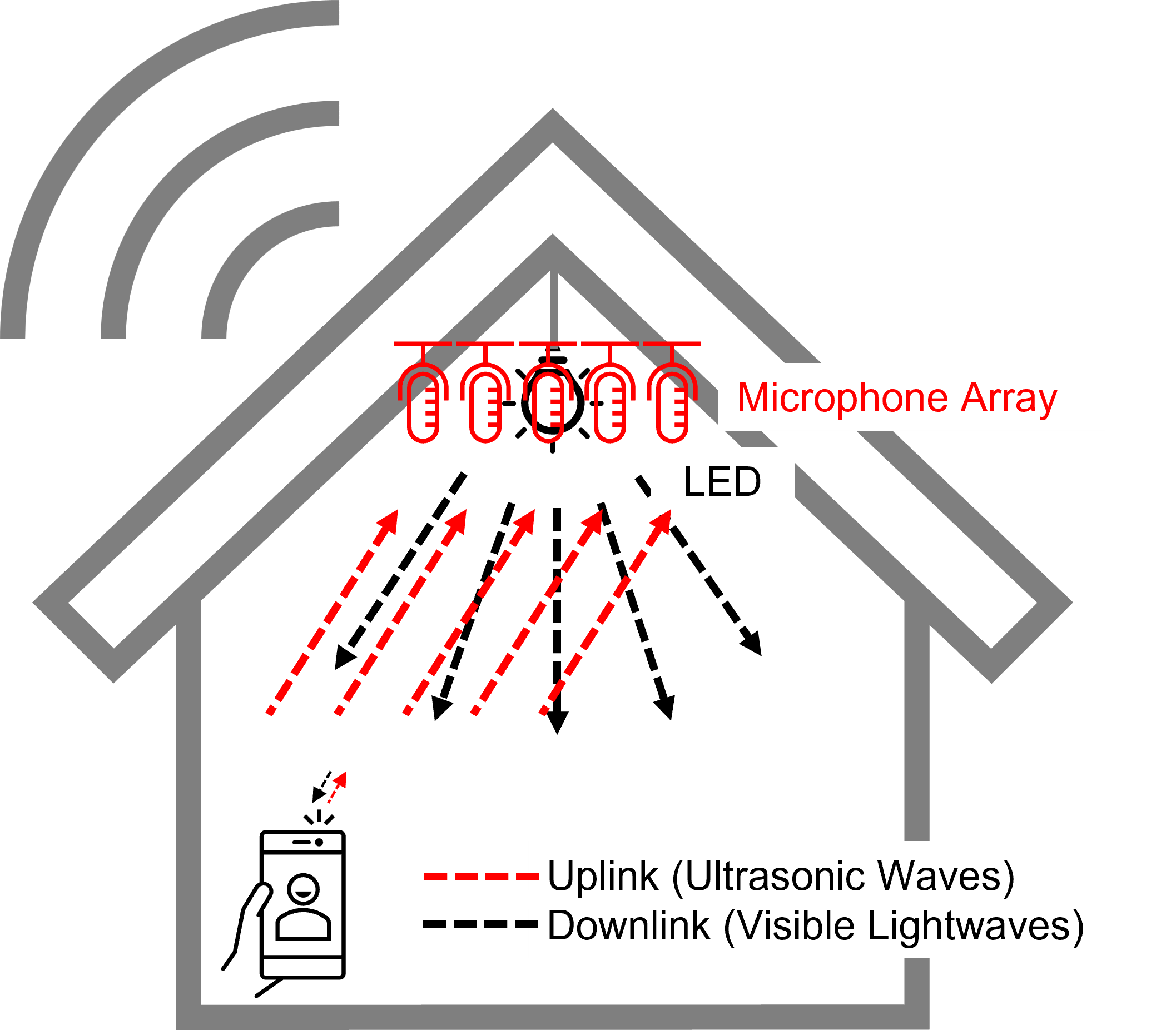}
\caption{Application scenarios using ultrasound for uplinking visible light communications.}
\label{scen}
\end{figure}

\begin{figure}[H]
\centering
\includegraphics[width=0.99\columnwidth]{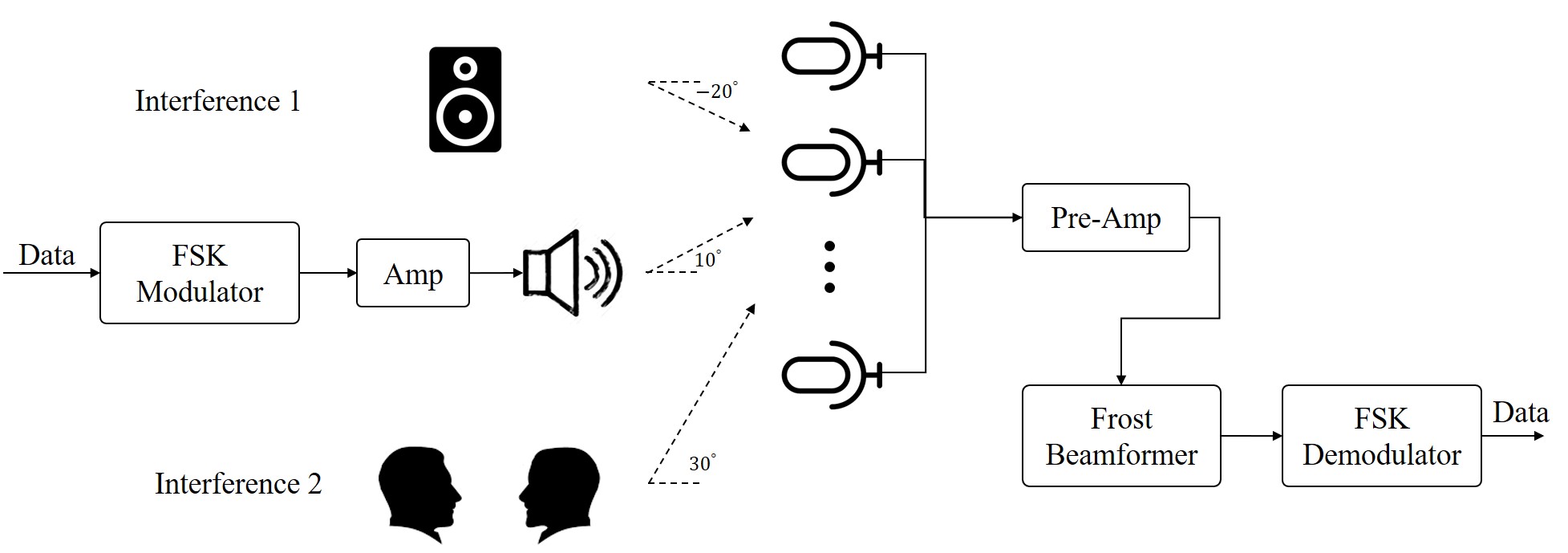}
\caption{The diagram on uplink transmission based on digital acoustic beam-forming using a microphone array.}
\label{chap3:uplink_audio}
\end{figure}

As shown in Figure \ref{scen}, this scheme is based on inaudible audio for uplink transmission of VLC. 
We propose a scheme that uses frequency-shift keying (FSK) to modulate data, and employs microphone array to process the received audio signal by digital beamforming algorithm. 
As illustrated in Figure~\ref{chap3:uplink_audio}, three audio signals come from three different directions, $-10^{\circ}$,  $-30^{\circ}$ and $20^{\circ}$. The linear microphone array consists of 10 omni-directional radiating microphones with 0.05-meter distance between adjacent microphones. Simulation results show that frost beamformer can effectively distinguish the uplink sound source.

\begin{figure}[H]
  \centering
  \subfigure[]{\includegraphics[width=0.9\textwidth]{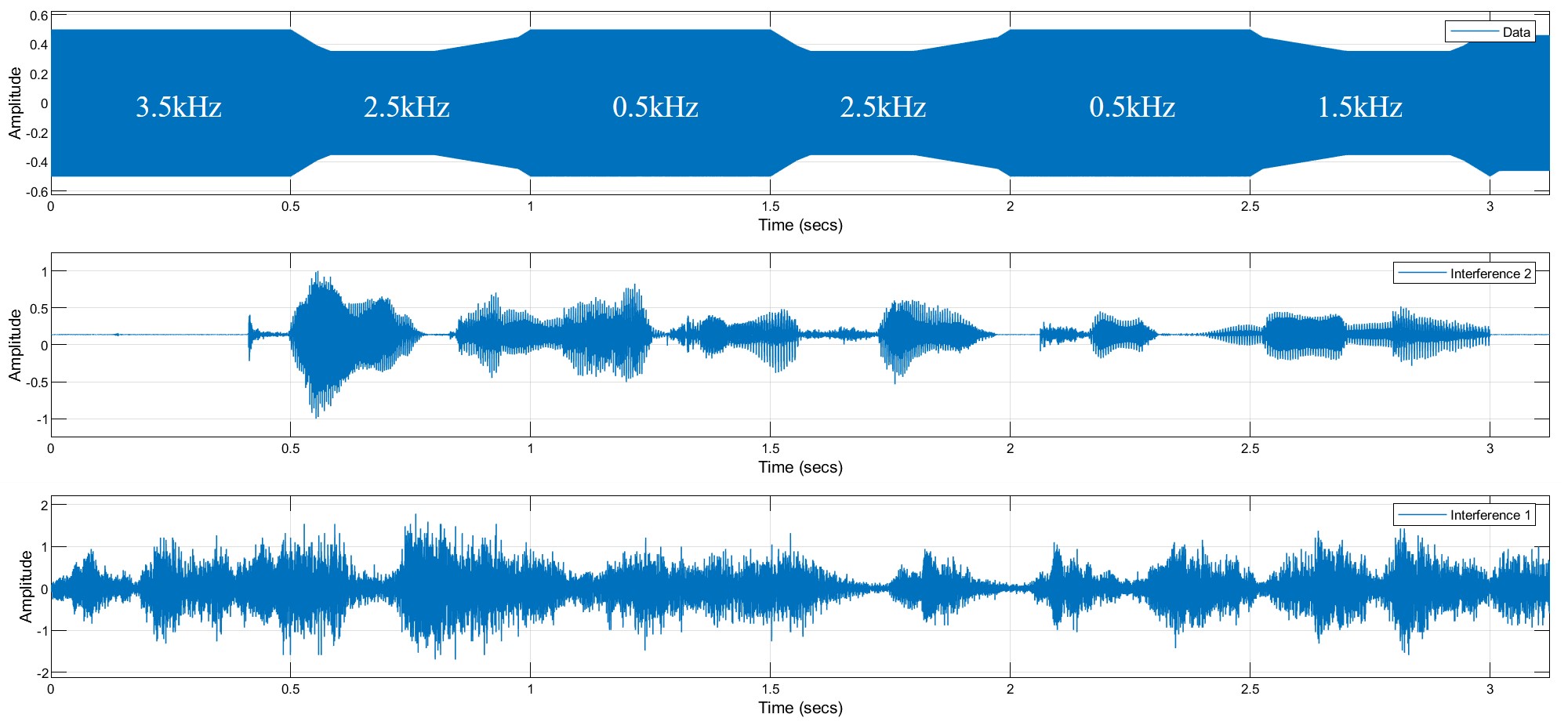}\label{chap3:uplink_wave1}}
  \subfigure[]{\includegraphics[width=0.9\textwidth]{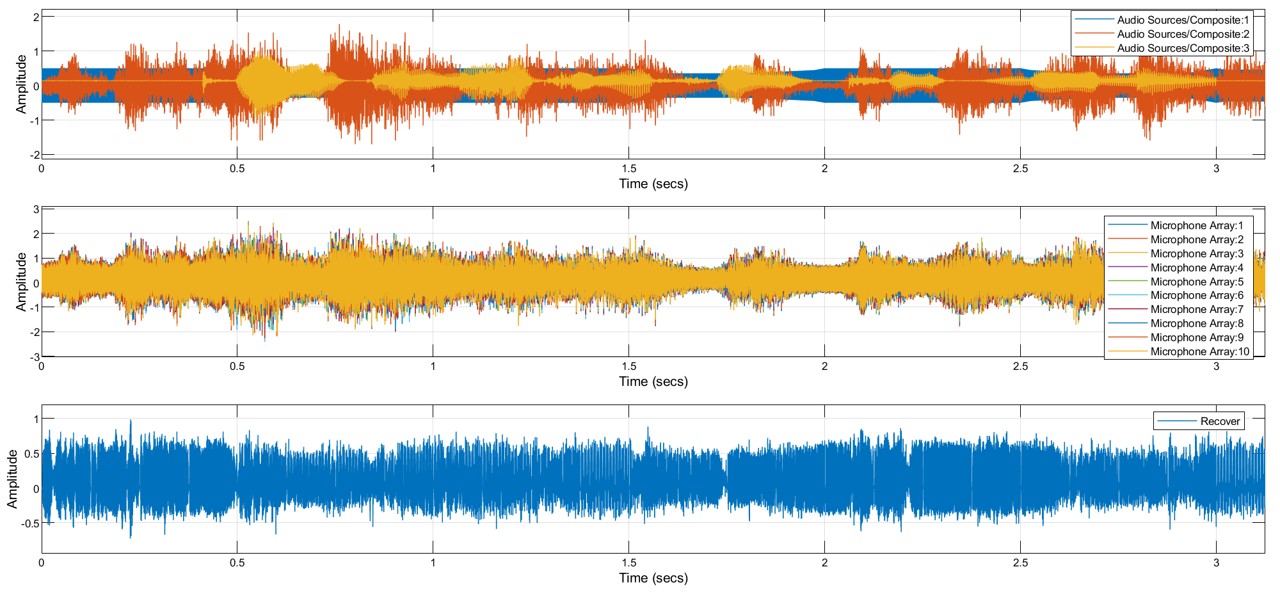}\label{chap3:uplink_wave2}}
  \caption{(a) The transmitted waveform of data, interference 1 and 2, (b) the composite waveform for audio source, the waveform received by each microphone, and recovered data signal by digital beamforming algorithm.}
  \label{chap3:uplink_wave1,uplink_wave2}
\end{figure}

According to tests, the frequency range of sound waves that can be heard by humans is 100 Hz to 15 kHz. For example, Apple's earphones have a response frequency from 5 Hz to 21 kHz \cite{apple_latin_america}.
This margin for inaudible frequencies can be used to transmit uplink data. 
To demonstrate this in a prototype verification experiment, four audible frequencies (0.5 kHz, 1.5 kHz, 2.5 kHz and 3.5 kHz) were used as carrier signals for 4-FSK, as seen in Figure~\ref{chap3:uplink_wave1}. 
The composite signal, which included data and interference signals, were received by 10 microphones and the data signal was recovered. The results show that the data signal was successfully recovered by beam-forming microphone arrays, as seen in Figure~\ref{chap3:uplink_wave2}.

Besides of capability without interference with downlink light, this technology can realize asymmetric uplink and downlink service bandwidth for Internet. Due to the users' internet habits, in most cases, the ratio of downlink bandwidth to uplink can reach a ratio of 10:1. Thus, it is reasonable for using this technique to solve the uplink scheme of visible light communication. 

\section{Conclusion}

This paper examines the difficulties encountered in accomplishing successful uplink transmission in VLC. Conventional techniques, such as physical reflectors or radio frequency technology, have their own drawbacks. To address this issue, an alternative scheme is proposed which employs inaudible audio signals for uplink transmission, with the aid of FSK modulation and a microphone array with beamforming algorithm to modulate and receive data, respectively. By taking advantage of the frequency margin which is not audible in general electronic product players, data transmission has been successfully achieved. This method not only eliminates interference with downlink light, but also provides asymmetric uplink and downlink service bandwidth, which is in line with user's internet habits.

\bibliographystyle{ieeetr} 
\bibliography{sample}

\begin{thebibliography}{10}

\bibitem{chap3_Komine_bidir}
T.~Komine, S.~Haruyama, and M.~Nakagawa, ``Bi-directional visible-light
  communication using corner cube modulator,'' in {\em Proceedings of the
  IASTED International Conference on Wireless and Optical Communications}
  (L.~Hesselink, ed.), vol.~3, pp.~598--603, 2003.

\bibitem{chap3_Shao_7035746}
S.~{Shao}, A.~{Khreishah}, M.~B. {Rahaim}, H.~{Elgala}, M.~{Ayyash}, T.~D.~C.
  {Little}, and J.~{Wu}, ``An indoor hybrid wifi-vlc internet access system,''
  in {\em 2014 IEEE 11th International Conference on Mobile Ad Hoc and Sensor
  Systems}, pp.~569--574, Oct 2014.

\bibitem{chap3_Shao_7293077}
S.~{Shao}, A.~{Khreishah}, M.~{Ayyash}, M.~B. {Rahaim}, H.~{Elgala},
  V.~{Jungnickel}, D.~{Schulz}, T.~D.~C. {Little}, J.~{Hilt}, and R.~{Freund},
  ``Design and analysis of a visible-light-communication enhanced wifi
  system,'' {\em IEEE/OSA Journal of Optical Communications and Networking},
  vol.~7, pp.~960--973, October 2015.

\bibitem{chap3_Yin_53643534}
S.~Yin, {\em Heterogeneous Wireless and Visible Light Communication for the
  Internet of Things}.
\newblock PhD thesis, University of Houston, 2018.

\bibitem{chap3_Kim_7423307}
M.~{Kim}, I.~{Jang}, S.~{Lim}, and T.~{Kang}, ``Implementation of zigbee-vlc
  system to support light control network configuration,'' in {\em 2016 18th
  International Conference on Advanced Communication Technology (ICACT)},
  pp.~1--1, Jan 2016.

\bibitem{chap3_Alresheedi_2017uplink}
M.~T. Alresheedi, A.~T. Hussein, and J.~M. Elmirghani, ``Uplink design in vlc
  systems with ir sources and beam steering,'' {\em IET Communications},
  vol.~11, no.~3, pp.~311--317, 2017.

\bibitem{chap3_Alsulami_2019infrared}
O.~Z. Alsulami, M.~T. Alresheedi, and J.~M. Elmirghani, ``Infrared uplink
  design for visible light communication (vlc) systems with beam steering,''
  {\em arXiv preprint arXiv:1904.02828}, 2019.

\bibitem{chap3_liu_2012demonstration}
Y.~Liu, C.~Yeh, C.~Chow, Y.~Liu, Y.~Liu, and H.~Tsang, ``Demonstration of
  bi-directional led visible light communication using tdd traffic with
  mitigation of reflection interference,'' {\em Optics express}, vol.~20,
  no.~21, pp.~23019--23024, 2012.

\bibitem{chap3_wang_2013875}
Y.~Wang, Y.~Shao, H.~Shang, X.~Lu, Y.~Wang, J.~Yu, and N.~Chi, ``875-mb/s
  asynchronous bi-directional 64qam-ofdm scm-wdm transmission over
  rgb-led-based visible light communication system,'' in {\em Optical Fiber
  Communication Conference}, pp.~OTh1G--3, Optical Society of America, 2013.

\bibitem{chap3_wang_2013demonstration}
Y.~Wang, Y.~Wang, N.~Chi, J.~Yu, and H.~Shang, ``Demonstration of 575-mb/s
  downlink and 225-mb/s uplink bi-directional scm-wdm visible light
  communication using rgb led and phosphor-based led,'' {\em Optics express},
  vol.~21, no.~1, pp.~1203--1208, 2013.

\bibitem{apple_latin_america}
Y.~A. Huang and J.~Benesty, {\em Audio signal processing for next-generation
  multimedia communication systems}.
\newblock Springer Science \& Business Media, 2007.

\end{thebibliography}

\end{document}